\documentclass[]{spie}  

\usepackage{amsmath,amsfonts,amssymb}
\usepackage{graphicx}
\usepackage[colorlinks=true, allcolors=blue]{hyperref}
\usepackage{adjustbox}
\usepackage{graphicx}
\usepackage{siunitx}
\usepackage{mathtools}
\usepackage{threeparttable}
\usepackage{float}
\usepackage{booktabs}
\usepackage{tabularx}
\usepackage{makecell}
\newcolumntype{C}[1]{>{\centering\arraybackslash}p{#1}}
\newcolumntype{L}[1]{>{\raggedright\arraybackslash}p{#1}}
\usepackage{multicol}
\usepackage[T1]{fontenc}
\usepackage{csquotes}
\usepackage{url}
\usepackage{placeins}

\usepackage{caption}
\usepackage{subcaption}
\usepackage{xcolor}
\usepackage{xr}
\usepackage{cleveref}
\usepackage{array}
\usepackage{multirow}
\usepackage{geometry}

\title{Leveraging Anatomical Priors for Automated \\ Pancreas Segmentation on Abdominal CT}

\author{Anisa V. Prasad}
\author{Tejas Sudharshan Mathai} 
\author{Pritam Mukherjee}
\author{Jianfei Liu}
\author{Ronald M. Summers}
\affil[]{Radiology and Imaging Sciences, National Institutes of Health Clinical Center, USA}

\authorinfo{Send correspondence to T.S.M.: tejas dot mathai at nih dot gov}

\begin{document} 
\maketitle

\begin{abstract}

An accurate segmentation of the pancreas on CT is crucial to identify pancreatic pathologies and extract imaging-based biomarkers. However, prior research on pancreas segmentation has primarily focused on modifying the segmentation model architecture or utilizing pre- and post-processing techniques. In this article, we investigate the utility of anatomical priors to enhance the segmentation performance of the pancreas. Two 3D full-resolution nnU-Net models were trained, one with 8 refined labels from the public PANORAMA dataset, and another that combined them with labels derived from the public TotalSegmentator (TS) tool. The addition of anatomical priors resulted in a 6\% increase in Dice score ($p < .001$) and a 36.5 mm decrease in Hausdorff distance for pancreas segmentation ($p < .001$). Moreover, the pancreas was always detected when anatomy priors were used, whereas there were 8 instances of failed detections without their use. The use of anatomy priors shows promise for pancreas segmentation and subsequent derivation of imaging biomarkers.

\end{abstract}


\section{Purpose}
\label{sec:Purpose}  

Imaging biomarkers have been extracted from the pancreas for early detection of diabetes \cite{Tallam2022,Schwartz2022,Suri2024}, incident pancreatic cancer \cite{Korfiatis2023,Alves2024,Cao2023,Mahmoudi2022,Toshima2021,Singh2020}, pancreatic cystic lesions \cite{Abel2021,Nader2023}, and opportunistic screening \cite{Sartoris2021}. Abdominal contrast-enhanced computed tomography (CECT) is the prevalent imaging technique for diagnosis of pancreatic cancer. An accurate segmentation of the pancreas is crucial for these applications as biomarkers obtained from an incorrectly segmented pancreas are unreliable \cite{Suri2024}.  However, it is difficult to segment on CT due to its small size, wide variability in morphology and appearance, and proximity to other structures of similar intensity (e.g., stomach) \cite{Suri2024,Salanitri2021,Chen2022}. The diverse CT scanner brands and exam protocols used at various institutions further render the task challenging.


There are numerous prior studies on pancreas segmentation \cite{Salanitri2021,Chen2022,Zhou2023_nnFormer,Tallam2022,Zhang2021,Korfiatis2023,Ji2022,Wasserthal2023_TS,Hatamizadeh2022,Ma2021}, but a majority of these approaches have modified the base segmentation model (U-Net), or used pre- and post-processing techniques for improved segmentation. Previous works \cite{Bagheri2020,Suri2024} have demonstrated that the presence of visceral adipose tissue around the pancreas can improve segmentation performance, but it may be reduced or absent in some patients. Anatomy priors, which are additional constraints used to guide segmentation models based on typical locations, shapes, and relationships of anatomical structures (e.g., kidney, colon), may help differentiate the pancreas better \cite{Dalca2018,Wasserthal2023_TS,Ma2021,Ji2022}. Their utility for pancreas segmentation, however, has not been directly studied.   

In this work, we solidify the utility of anatomy priors and propose a fully automated deep learning-based pipeline that uses them directly to segment the pancreas. The public PANORAMA dataset \cite{Alves2024} was used for training 3D nnU-Net models with and without priors. The models were evaluated on the public AMOS22 dataset \cite{Ji2022}. The anatomy priors incorporated to segment the pancreas are promising for the automated computation of pancreatic imaging biomarkers.

\begin{figure} [!t]
\begin{center}
\begin{tabular}{c} 
\includegraphics[page=9, width=0.95\textwidth]{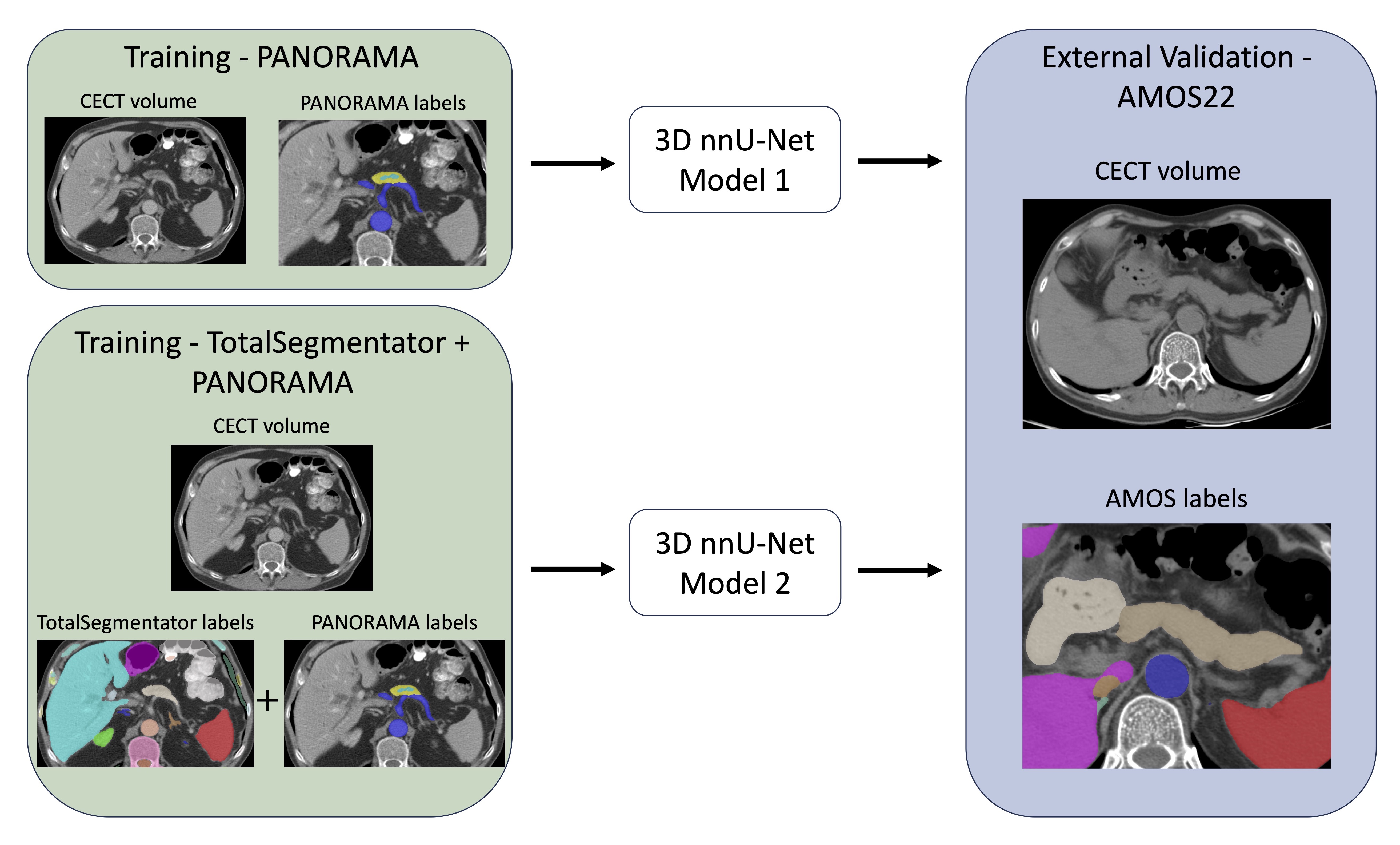}
\end{tabular}
\end{center}
\caption{Fully automated framework for pancreas segmentation on CT. Two 3D nnU-Net models were trained with and without anatomy priors. Model 1 (top) was trained with refined PANORAMA dataset labels, while Model 2 (bottom) was trained with the refined labels and anatomy priors (constraints used to guide segmentation based on anatomical structures) obtained from the public TotalSegmentator (TS) tool. Both models were evaluated on the AMOS22 dataset.}
\label{fig_money} 
\end{figure} 

\section{METHODS}
\label{sec_methods}

\begin{figure} [H]
\begin{center}
\begin{tabular}{c} 
\includegraphics[page=3, width=0.95\textwidth]
{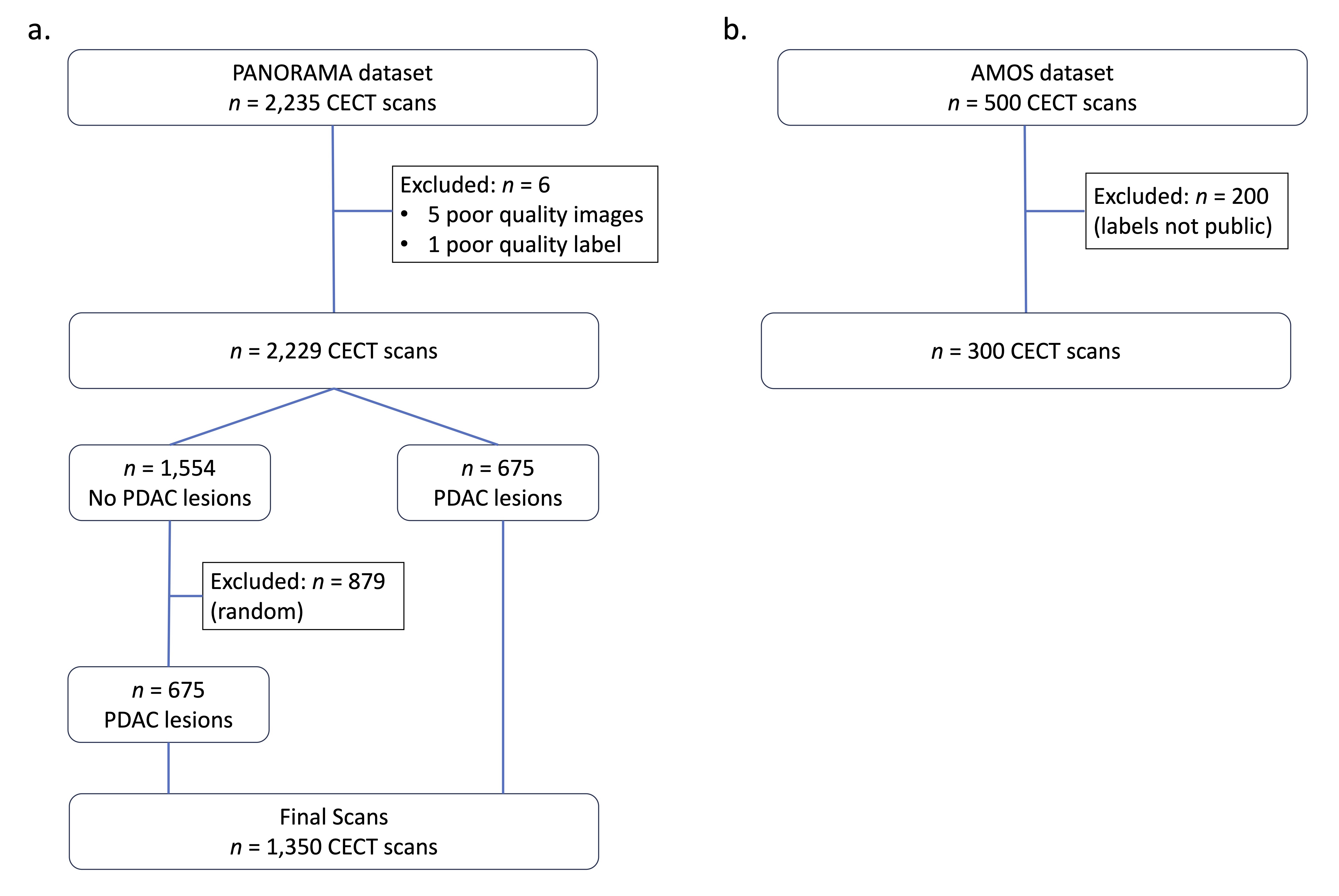}
\end{tabular}
\end{center}
\caption[example] 
{ \label{fig:STARD} 
Standards for Reporting Diagnostic Accuracy (STARD) Chart describing the inclusion and exclusion criteria of patients for the (a) PANORAMA dataset (training) and (b) external AMOS22 dataset (testing). }
\end{figure}

\noindent
\textbf{Patient Sample.} In this retrospective study, two public datasets containing abdominal CT scans were used. The public PANORAMA dataset contains 2,235 portal-venous CT scans and corresponding ground truth labels of pancreatic anatomy \cite{Alves2024}. 6 scans were excluded due to corruption, which left a total of 2,229 CT scans containing an unequal mix of PDAC and non-pancreatic ductal adenocarcinoma (PDAC) patients. The dataset was balanced to ensure an equal distribution of PDAC and non-PDAC patients (n = 675), resulting in 1,350 CT volumes for training. The public AMOS22 dataset was used for testing. It contains 500 CECT scans, 300 of which have been publicly released. The ground-truth labels (pancreas and 14 other organs) for these volumes were available \cite{Ji2022}. Supplemental Fig. \ref{fig:STARD} summarizes the data collection process.

\noindent
\textbf{Reference Standard.} All the arteries feeding the pancreas (including the aorta) were annotated with one label in the PANORAMA dataset. To maintain consistency with the AMOS dataset labels used for testing, which contains just an aorta label, the PANORAMA arteries label was divided into the aorta and the remaining arteries. The public TotalSegmentator (TS) tool \cite{Wasserthal2023_TS}, which generated segmentations of 117 structures, was used to do this. The TS aorta label was masked out of the PANORAMA arteries label, leaving only the pancreatic arteries (superior mesenteric artery, splenic artery, and pancreatico-duodenal artery). Through this process, the original PANORAMA labels were refined into 8 labels: (1) background, (2) pancreatic duct, (3) common bile duct, (4) aorta, (5) PDAC lesion, (6) veins, (7) arteries (excluding aorta), and (8) pancreas. These refined 8 labels were used for training a model. Next, the TS labels were combined with the refined labels to yield a total of 45 labels. All 117 structures segmented by TS are represented in these 45 labels, as some were combined into a single label to simplify the segmentation task. In case of an overlap between TS and PANORAMA (e.g., pancreas), the refined PANORAMA label was conserved. These 45 labels were used to train another model. See Supplemental Table \ref{table_anatomy_priors_def} for further details.

\noindent
\textbf{Model.} Fig. \ref{fig_money} shows the framework overview. The self-configuring nnU-Net framework \cite{Isensee2021} was used as it is the \textit{de-facto} standard for segmentation tasks due to its award-winning performance \cite{Isensee2021,Isensee2024} and rigorous validation \cite{Isensee2024} for many tasks, including multi-organ segmentation in CT and MRI \cite{Zhuang2024,Wasserthal2023_TS} among others. It has often outperformed other architectures \cite{Isensee2024}, such as transformer-based approaches \cite{Zhou2023_nnFormer}. The framework automatically determines a ``fingerprint'' for the input CT volumes that includes intensity normalization and resampling to a consistent spacing among others. The model learned to segment the targets in the CT volume and iteratively refined it via a loss function, which was a combination of the Dice loss and the binary cross entropy loss \cite{drozdzal2016importance}. Two 3D full-resolution nnU-Net models were trained: (1) ``\texttt{REF$\_$8}'' with the 8 refined PANORAMA labels, and (2) ``\texttt{ALL$\_$45}'' with the 8 refined and 37 anatomy priors from TS (total 45 labels). Additional implementation details are provided in the Appendix.

\noindent
\textbf{Statistical Analysis.} Metrics, such as Dice similarity coefficient (DSC) and Hausdorff Distance (HD), were computed. Any failure to detect and segment the pancreas was identified. Friedman tests (for DSC and HD) and Cochran's Q tests (detection failure) were performed, and a $p < .05$ was considered statistically significant. Post-hoc analyses included Nemenyi tests (DSC and HD) and McNemar's tests with manual Bonferroni corrections (detection failure).

\section{Results}
\label{sec_results}

Table \ref{tab_organ_metrics} describes quantitative pancreas segmentation performance of TS and both models. Qualitative results for each model can be found in Supplemental Fig. \ref{panc_seg}. The model trained with all 45 anatomy prior labels (\texttt{ALL$\_$45}) segmented the pancreas better (DSC: 0.81 $\pm$ 0.14, HD: 24.0 $\pm$ 65.0, $p < .001$) compared to \texttt{REF$\_$8} (DSC: 0.75 $\pm$ 0.22, HD: 60.5 $\pm$ 150.2). The differences were significant for both DSC and HD error (Table \ref{tab_pvalues}). The detection failure rate also markedly dropped from 8 missed detections of the pancreas with \texttt{REF$\_$8} to 0 missed detections with \texttt{ALL$\_$45}. The DSC from TS alone (0.80 $\pm$ 0.13) was higher than the model without priors, but lower than the model with priors. It had the lowest HD error across all 3 models (21.5 $\pm$ 71.1), and 0 detection failures. From Table \ref{tab_pvalues}, there was a difference ($p = .001$) between TS and ALL-45 in terms of DSC and HD errors. Similarly, there was a difference between ALL-45 and REF-8 ($p < .05$) across all metrics.   

\begin{table}[!htb]
\centering
\caption{Pancreas segmentation performance by TS and the models trained with and without anatomical priors. Bold font indicates best results. Friedman and Cochran's Q tests yielded p-values, which indicated any statistical difference between the 3 models.}
\smallskip
\begin{adjustbox}{max width=0.95\textwidth}
\begin{tabular}{@{} l *{4}{c} @{}}
  \toprule
  Metric & TS & REF-8 & ALL-45 & p-value \\
  \midrule
  DSC & 0.80 $\pm$ 0.13 &  0.75 $\pm$ 0.22 & \textbf{0.81 $\pm$ 0.14} & $<$ 0.001 \\
  HD (mm) & \textbf{21.5 $\pm$ 71.1} & 60.5 $\pm$ 150.2 & 24.0 $\pm$ 65.0 & $<$ 0.001 \\
  Detection failure (n) & 0 & 8 & 0 & $<$ 0.001 \\
  \bottomrule
\end{tabular}
\end{adjustbox}
\label{tab_organ_metrics}
\end{table}
\begin{table}[!htb]
\centering
\caption{Pairwise comparisons of model performance using a post-hoc analysis. Bold font indicates significant p-values.}
\smallskip
\begin{adjustbox}{max width=0.95\textwidth}
\begin{tabular}{@{} l *{3}{c} @{}}
  \toprule
  Metric & TS vs. REF-8 & TS vs. ALL-45 & REF-8 vs. ALL-45 \\
  \midrule
  DSC & 0.700 &  \textbf{0.001} & \textbf{0.001} \\
  HD (mm) & \textbf{0.039} & \textbf{0.001} & \textbf{0.001} \\
  Detection failure (n) & \textbf{0.023} & 1.000 & \textbf{0.023} \\
  \bottomrule
\end{tabular}
\end{adjustbox}
\label{tab_pvalues}
\end{table}

\section{Discussion}
\label{sec_discussion}

The pancreas is a small organ that is in close proximity to other structures (e.g., stomach, kidney). It heterogeneously enhances in CT, depending on the amount of intrapancreatic fat. The surrounding attenuation of the visceral organs (especially the pancreas) can affect the segmentation performance \cite{Suri2024}. Due to this, the accuracy of automated pancreas segmentation can depend on several factors, such as the imaging modality, quality and characteristics of the training dataset, model architecture, and pre- and post-processing techniques \cite{Bagheri2020}. Anatomical priors have been effective  \cite{Zhang2021,Dalca2018,Zhu2023,Ravishankar2017,Wang2023,Mathai2024,Sheng2024} at improving segmentation of anatomical structures. These shape priors can help resolve ambiguities that arise due to CT contrast status \cite{Suri2024}, low attenuation, and scan artifacts \cite{Bagheri2020}. 

In this work, anatomical priors from TS were leveraged to boost the pancreas segmentation performance with a 3D full-resolution nnU-Net. Our results are similar to those obtained by prior work \cite{Suri2024}, in which several models, such as Abdomen-Atlas UNet, Abdomen-Atlas Swin, MSD-nnU-Net, and DM-UNet, were used for pancreas segmentation and a maximum Dice score of 0.79 was obtained.  These results underscore the complexity surrounding pancreas segmentation. The variability in metrics decreased with the use of anatomical priors: the standard deviations of DSC and HD were notably lower in the model trained with priors. The extra labels penalized the model when it encroached into surrounding anatomy, resulting in more precise pancreas segmentations.  

The trained models label the vasculature and bile ducts, which traverse the pancreas. Both TS and the AMOS22 dataset, however, simply label the entire pancreas without distinguishing these internal structures. As a result, the TS pancreas label is inherently more aligned with the AMOS22 ground truth, as neither identifies any internal pancreatic anatomy such as the vasculature. This may contribute to underestimation of the trained models' pancreas segmentation accuracy.

Another limitation of our work is that the models were trained on CECT scans, and their performance on non-contrast CT scans is unclear. Both models were trained on the PANORAMA dataset containing portal-venous phase CT scans, and evaluated on the external AMOS22 dataset containing CECT scans. It is known that the performance of a model trained on CECT for pancreas segmentation drops on non-contrast CT \cite{Tallam2022} due to the ambiguous boundaries between adjacent anatomy. The exploration of this topic is left for future research. 

In summary, anatomical priors can constrain deep learning models to correctly segment the pancreas. This may be useful for automated extraction of CT-derived biomarkers for early diagnosis of pancreatic cancer and diabetes.

\clearpage 

\section{Acknowledgements}

This work was supported by the Intramural Research Program of the National Institutes of Health (NIH) Clinical Center (project number 1Z01 CL040004). The research used the high-performance computing facilities of the NIH Biowulf cluster.

\bibliography{references} 
\bibliographystyle{spiebib} 

\clearpage

\section{Appendix}
\label{sec_appendix}

\subsection{Model Details}

The base network is a UNet \cite{Isensee2021}, which was automatically optimized for the dataset based on its fingerprint. This included the determination of optimal hyper-parameters, such as a large batch size, adequate network depth among others. Other parameters, such as the number of epochs for training, were set to default values for training the 3D full-resolution nnU-Net model. To optimize the network weights, stochastic gradient descent with Nesterov momentum ($\mu$ = 0.99), Leaky ReLU nonlinear activation, an initial learning rate of $10^{-2}$, and a batch size of 2 were utilized. For each output from the model, the corresponding ground truth segmentation mask was used for calculating the losses. A variety of data augmentations (e.g., flipping, rotation, zoom) were implemented following the original implementation \cite{Isensee2021}. All experiments were done on the NIH High Performance Computing Cluster using either an NVIDIA Tesla V100x or A100 GPU. Experiments to determine the right loss function to use were unnecessary as it was empirically found in previous investigations \cite{Isensee2021,drozdzal2016importance} that the combination of the Dice and cross-entropy losses improved training stability and segmentation accuracy. 



\begin{table*}[t]
\centering\fontsize{9}{12}\selectfont 
\setlength\aboverulesep{0pt}\setlength\belowrulesep{0pt} 
\setlength{\tabcolsep}{7pt} 
\setcellgapes{3pt}\makegapedcells 
\caption{Complete list of all structures used for training the 3D full-resolution nnU-Net models in this work. The PANORAMA dataset provided labels for 6 structures (labels 38-39 and labels 41-44), while the labels for the remaining 38 structures were generated by TS (labels 1-37 and 40). The aorta in the PANORAMA arteries label (43) was masked out using the aorta label from TS (40). Check marks indicate the structures that were used to train each model.}
\begin{adjustbox}{max width=\textwidth}
\begin{tabular}{@{} |c|c|c|c|c| @{}} 
\toprule

\#      &   Structure                           & Model 1 (\texttt{REF$\_$8}) & Model 2 (\texttt{ALL$\_$45}) & Extracted From \\
\midrule

0       &   Background                                      & \checkmark              & \checkmark     & -  \\
1       &   Body region mask                                &               & \checkmark     & TS  \\
2       &   Spleen                                          &               & \checkmark     & TS \\
3       &   Kidneys (left \& right)                         &               & \checkmark     & TS \\
4       &   Gallbladder                                     &               & \checkmark     & TS \\
5       &   Liver                                           &               & \checkmark     & TS \\
6       &   Stomach                                         &               & \checkmark     & TS \\
7       &   Adrenal glands (left \& right)                  &               & \checkmark     & TS \\
8       &   Lungs (all lobes)                               &               & \checkmark     & TS \\
9       &   Esophagus                                       &               & \checkmark     & TS  \\
10      &   Trachea                                         &               & \checkmark     & TS \\
11      &   Thyroid                                         &               & \checkmark     & TS \\
12      &   Small bowel                                     &               & \checkmark     & TS \\
13      &   Duodenum                                        &               & \checkmark     & TS \\
14      &   Colon                                           &               & \checkmark     & TS \\
15      &   Urinary bladder                                 &               & \checkmark     & TS \\
16      &   Prostate                                        &               & \checkmark     & TS \\
17      &   Kidney cysts (left \& right)                    &               & \checkmark     & TS \\
18      &   \shortstack{Skeleton (vertebrae \& ribs \& skull \\ \& sacrum \& humerus \& scapula \\ \& clavicula \& femur \& hip \& sternum)}                                                                                                   &               & \checkmark     & TS \\
19      &   Heart \& left atrial appendage                  &               & \checkmark     & TS \\
20      &   Pulmonary vein                                  &               & \checkmark     & TS \\
21      &   Brachiocephalic trunk                           &               & \checkmark     & TS \\
22      &   Subclavian artery (left \& right)               &               & \checkmark     & TS \\
23      &   Common carotid artery                           &               & \checkmark     & TS \\
24      &   Brachiocephalic vein (left \& right)            &               & \checkmark     & TS \\
25      &   Superior vena cava                              &               & \checkmark     & TS \\
26      &   Inferior vena cava                              &               & \checkmark     & TS \\
27      &   Portal \& splenic vein                          &               & \checkmark     & TS \\
28      &   Common iliac artery (left \& right)             &               & \checkmark     & TS \\
29      &   Common iliac vein (left \& right)               &               & \checkmark     & TS \\
30      &   Spinal cord                                     &               & \checkmark     & TS \\
31      &   Gluteus maximus (left \& right)                 &               & \checkmark     & TS \\
32      &   Gluteus medius (left \& right)                  &               & \checkmark     & TS \\
33      &   Gluteus minimus (left \& right)                 &               & \checkmark     & TS \\
34      &   Autochthon                                      &               & \checkmark     & TS \\
35      &   Iliopsoas                                       &               & \checkmark     & TS \\
36      &   Brain                                           &               & \checkmark     & TS \\
37      &   Costal cartilage                                &               & \checkmark     & TS \\
38      &   Pancreatic duct                                 & \checkmark    & \checkmark     & PANORAMA \\
39      &   Common bile duct                                & \checkmark    & \checkmark     & PANORAMA \\
40      &   Aorta                                           & \checkmark    & \checkmark     & TS \\
41      &   PDAC lesion                                     & \checkmark    & \checkmark     & PANORAMA \\
42      &   Veins                                           & \checkmark    & \checkmark     & PANORAMA \\
43      &   Arteries excluding aorta                        & \checkmark    & \checkmark     & PANORAMA modified using TS \\
44      &   Pancreas parenchyma                              & \checkmark    & \checkmark     & PANORAMA \\

\bottomrule
\end{tabular}
\end{adjustbox}
\label{table_anatomy_priors_def}
\end{table*}

\begin{figure}[b]
    \centering
    \begin{subfigure}[b]{0.30\textwidth}
        \centering
        \includegraphics[width=\textwidth]{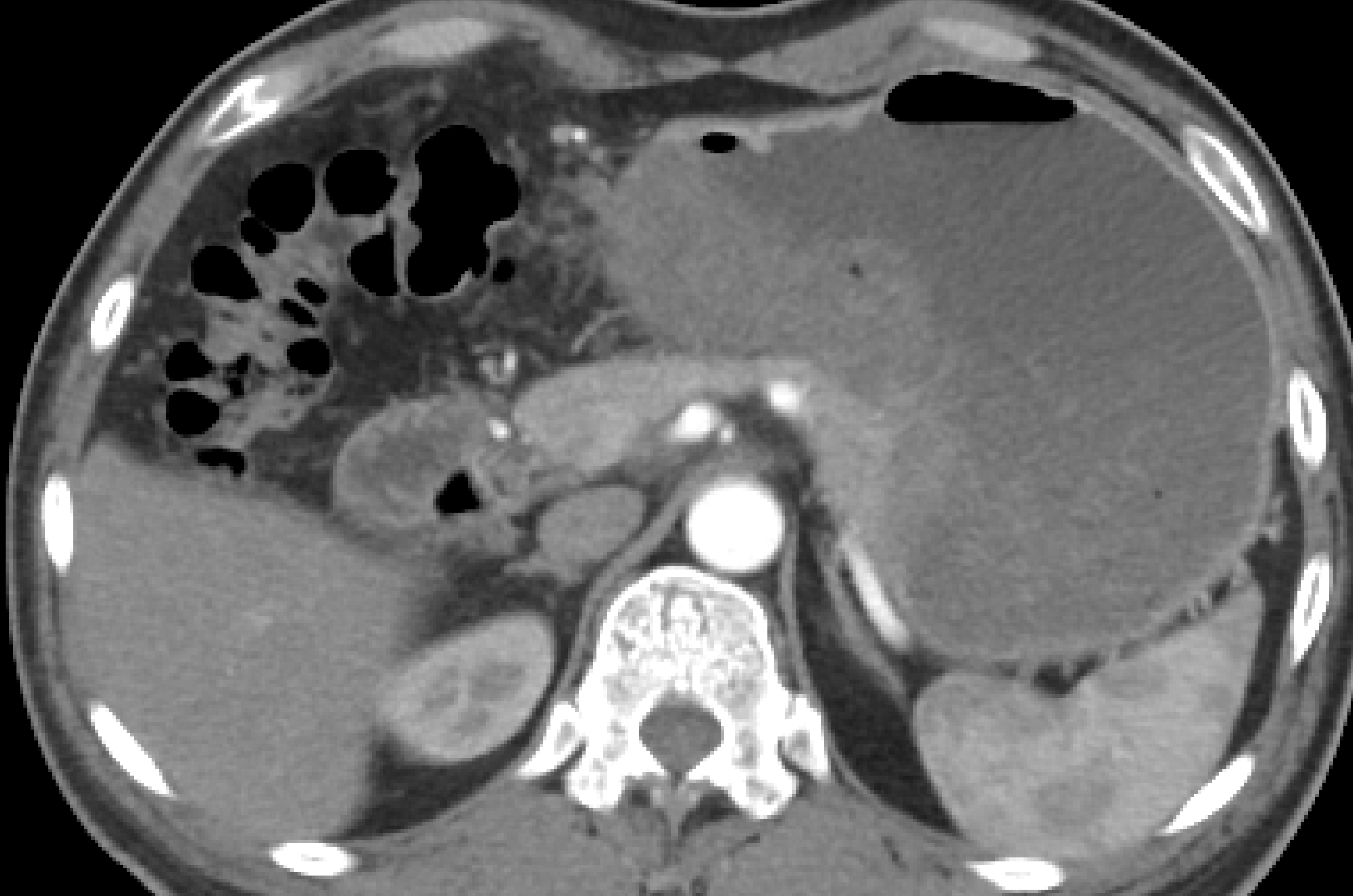}
        \caption{\small Case 1: CECT scan}
    \end{subfigure}
    \hspace{0.1\textwidth}
    \begin{subfigure}[b]{0.30\textwidth}
        \centering
        \includegraphics[width=\textwidth]{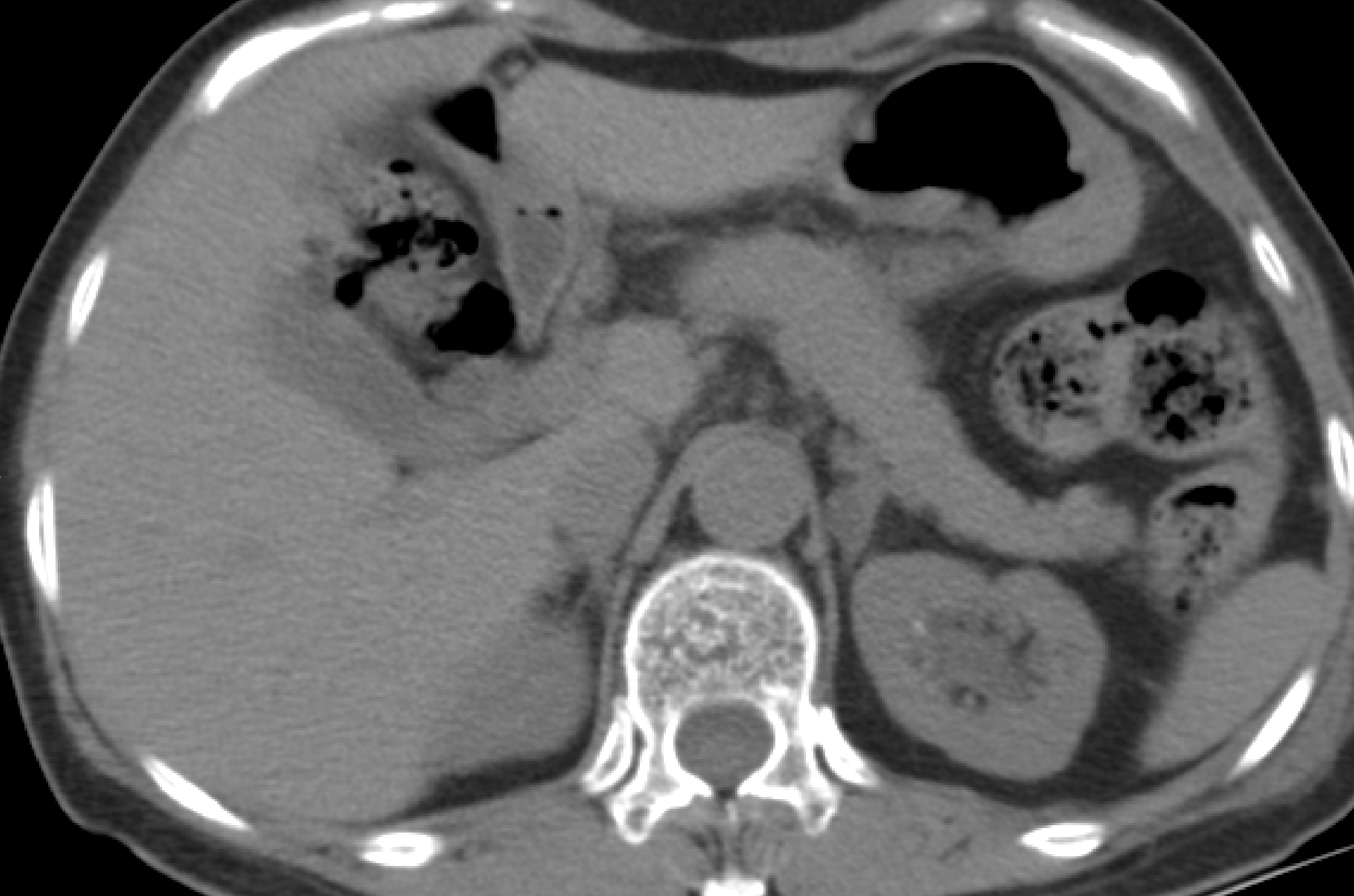}
        \caption{\small Case 2: CECT scan}
    \end{subfigure}
    \vskip\baselineskip
    \begin{subfigure}[b]{0.30\textwidth}  
        \centering 
        \includegraphics[width=\textwidth]{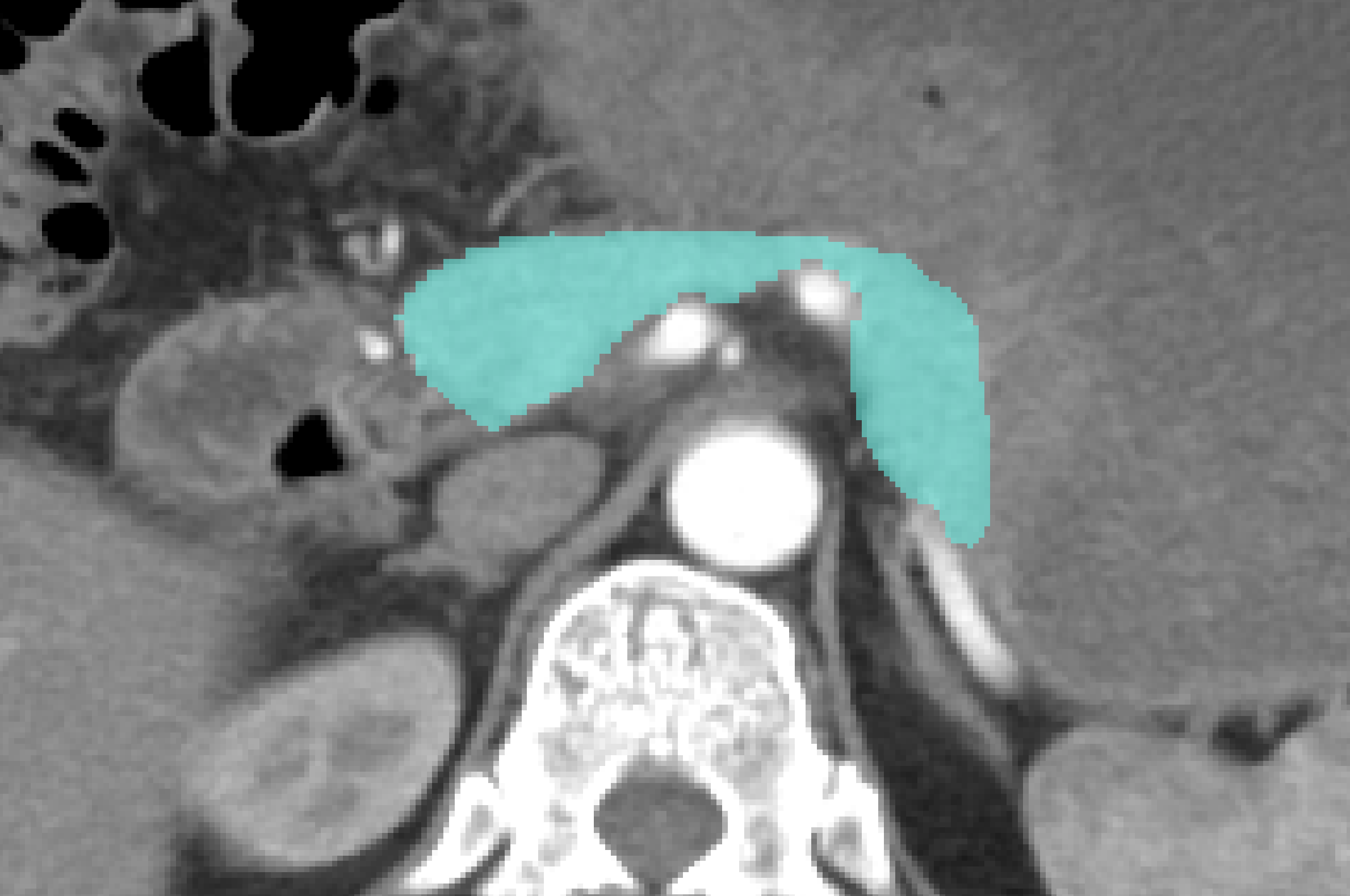}
        \caption{\small Case 1: Reference}
    \end{subfigure}
    \hspace{0.1\textwidth} 
    \begin{subfigure}[b]{0.30\textwidth}  
        \centering 
        \includegraphics[width=\textwidth]{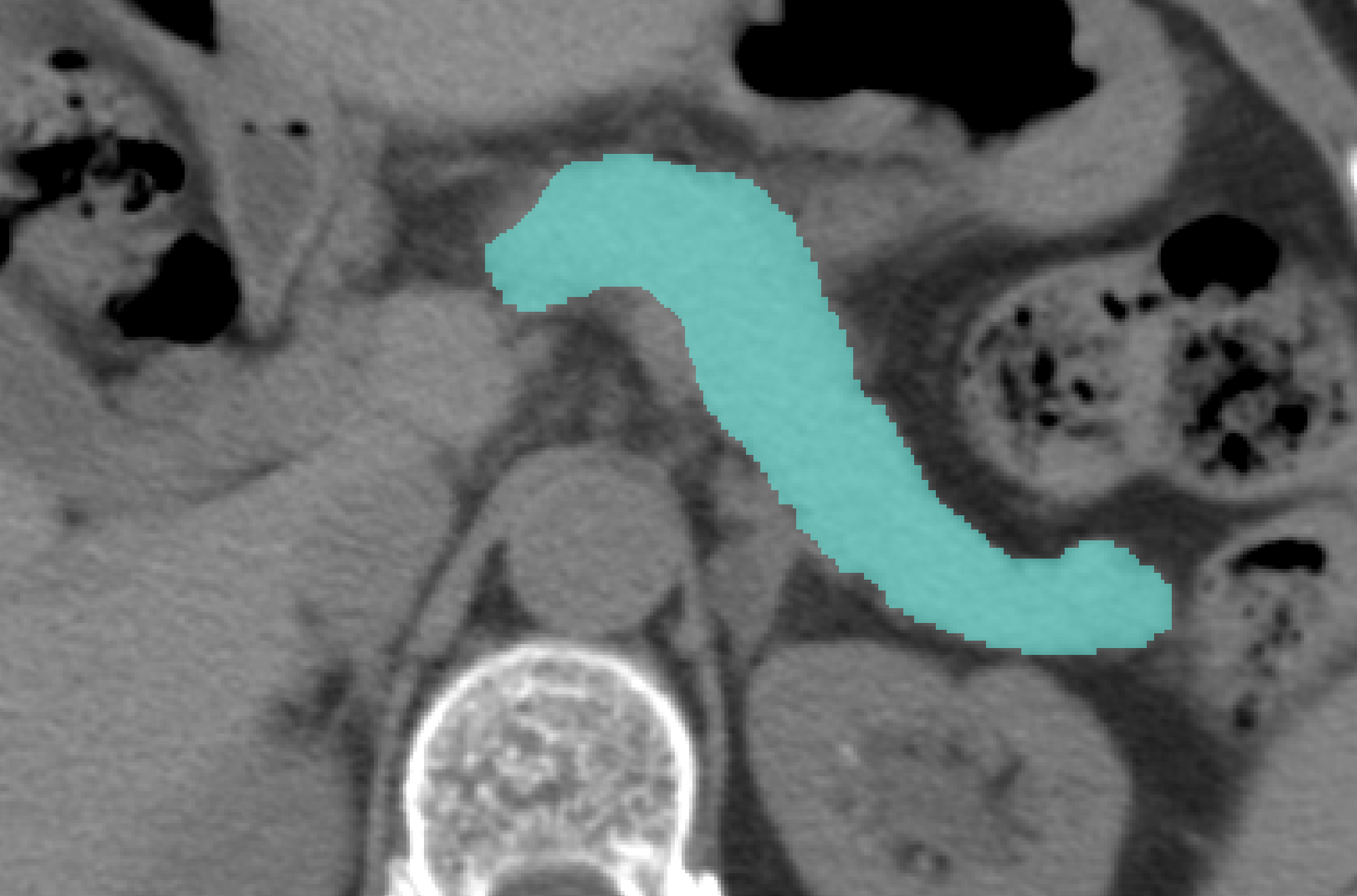}
        \caption{\small Case 2: Reference}
    \end{subfigure}
    \vskip\baselineskip
    \begin{subfigure}[b]{0.30\textwidth}   
        \centering 
        \includegraphics[width=\textwidth]{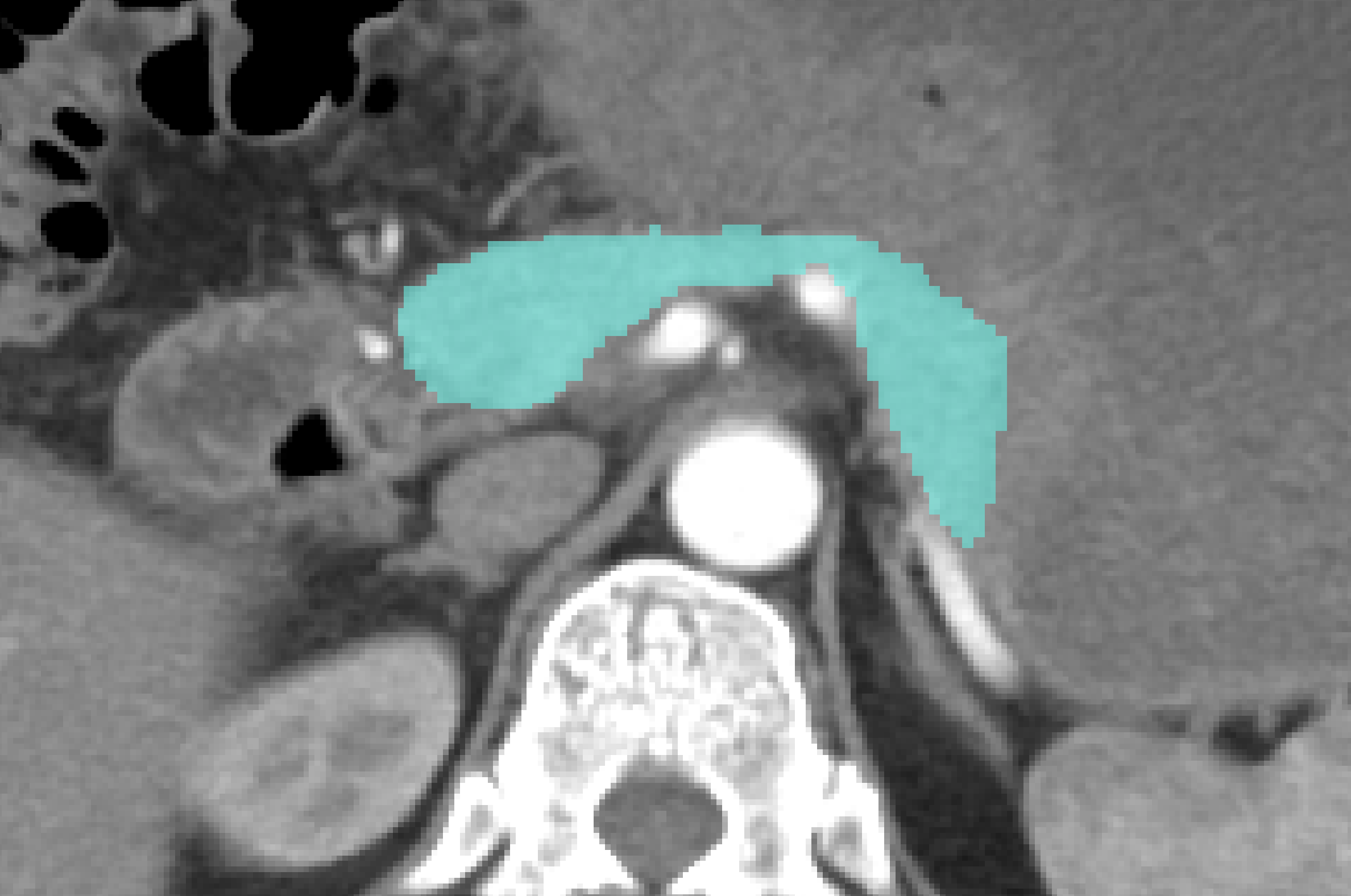}
        \caption{\small Case 1: TS Segmentation}
    \end{subfigure}
    \hspace{0.1\textwidth}
    \begin{subfigure}[b]{0.30\textwidth}   
        \centering 
        \includegraphics[width=\textwidth]{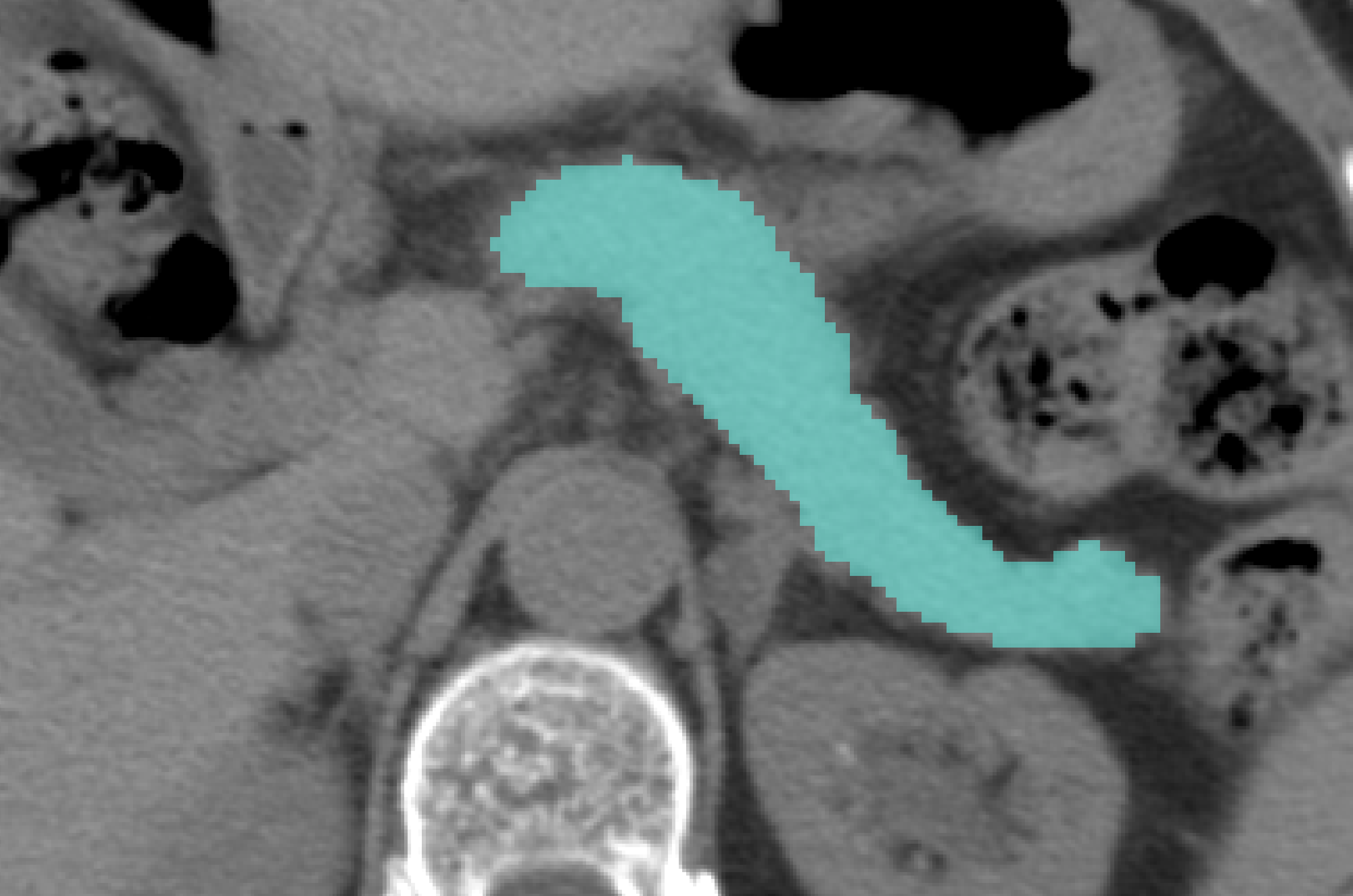}
        \caption{\small Case 2: TS Segmentation}
    \end{subfigure}
    \vskip\baselineskip
    \begin{subfigure}[b]{0.30\textwidth}   
        \centering 
        \includegraphics[width=\textwidth]{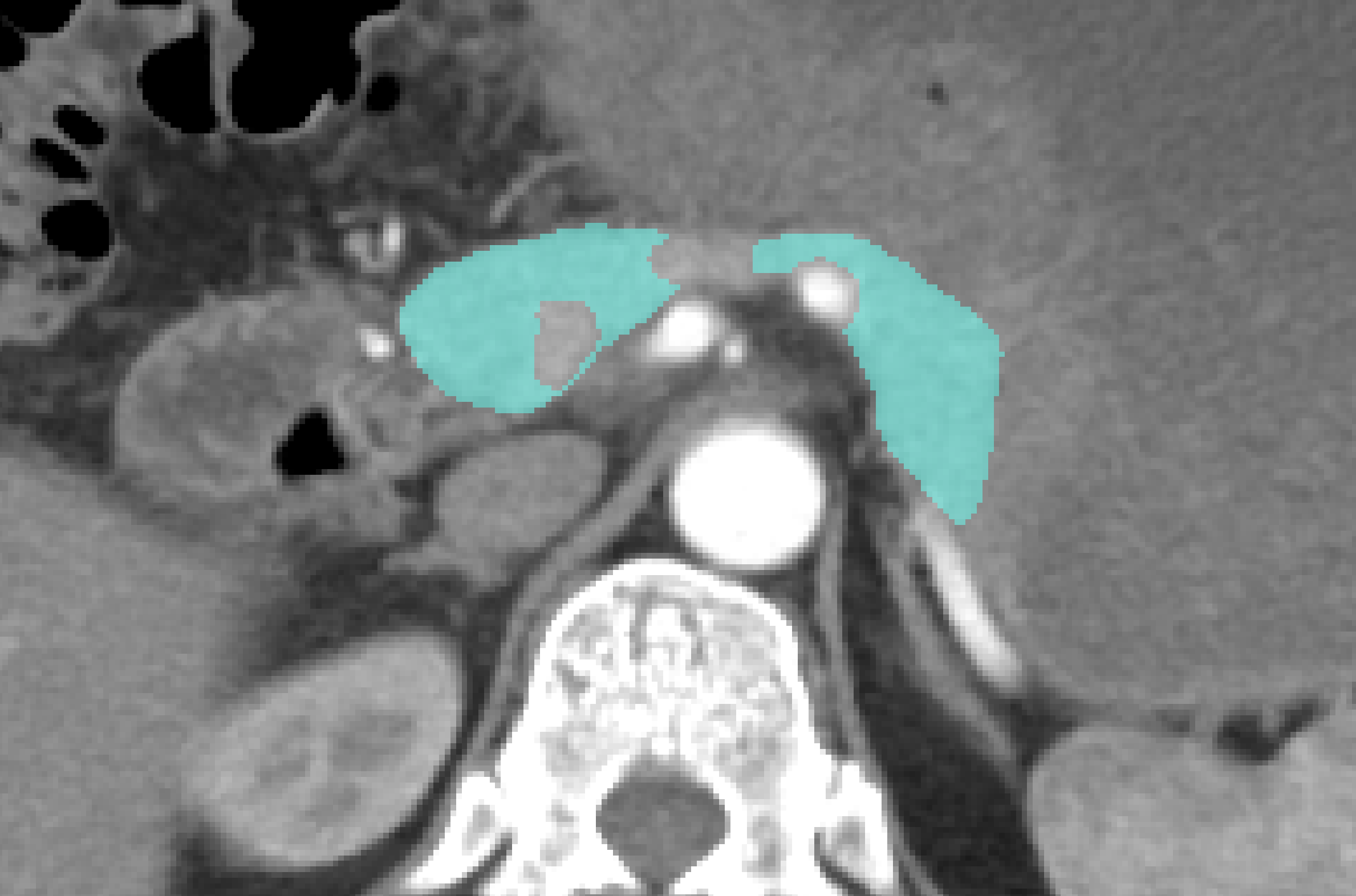}
        \caption{\small Case 1: Model 1 (\texttt{REF$\_$8})}
    \end{subfigure}
    \hspace{0.1\textwidth}
    \begin{subfigure}[b]{0.30\textwidth}   
        \centering 
        \includegraphics[width=\textwidth]{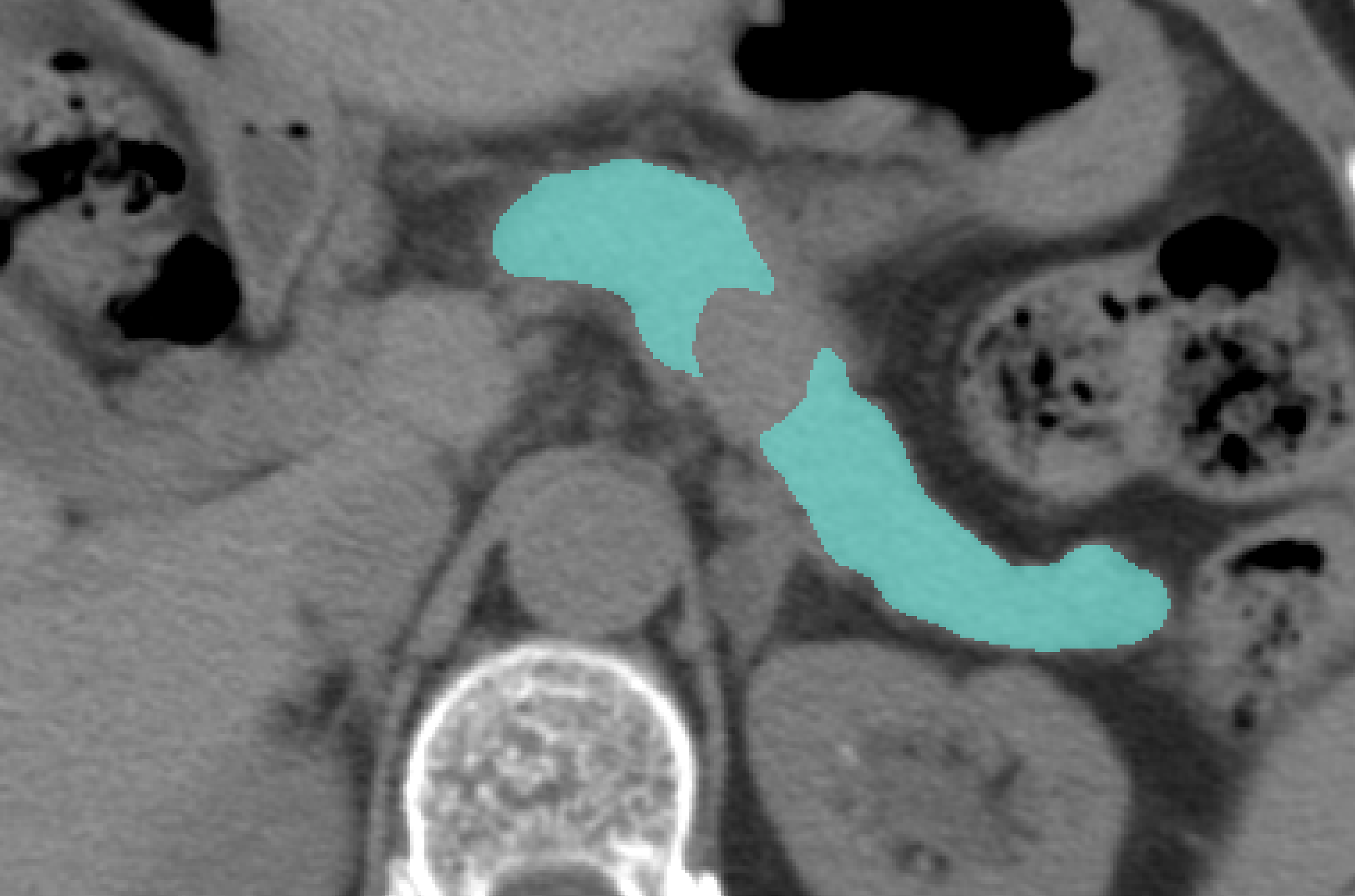}
        \caption{\small Case 2: Model 1 (\texttt{REF$\_$8})}
    \end{subfigure}
    \vskip\baselineskip
    \begin{subfigure}[b]{0.30\textwidth}   
        \centering 
        \includegraphics[width=\textwidth]{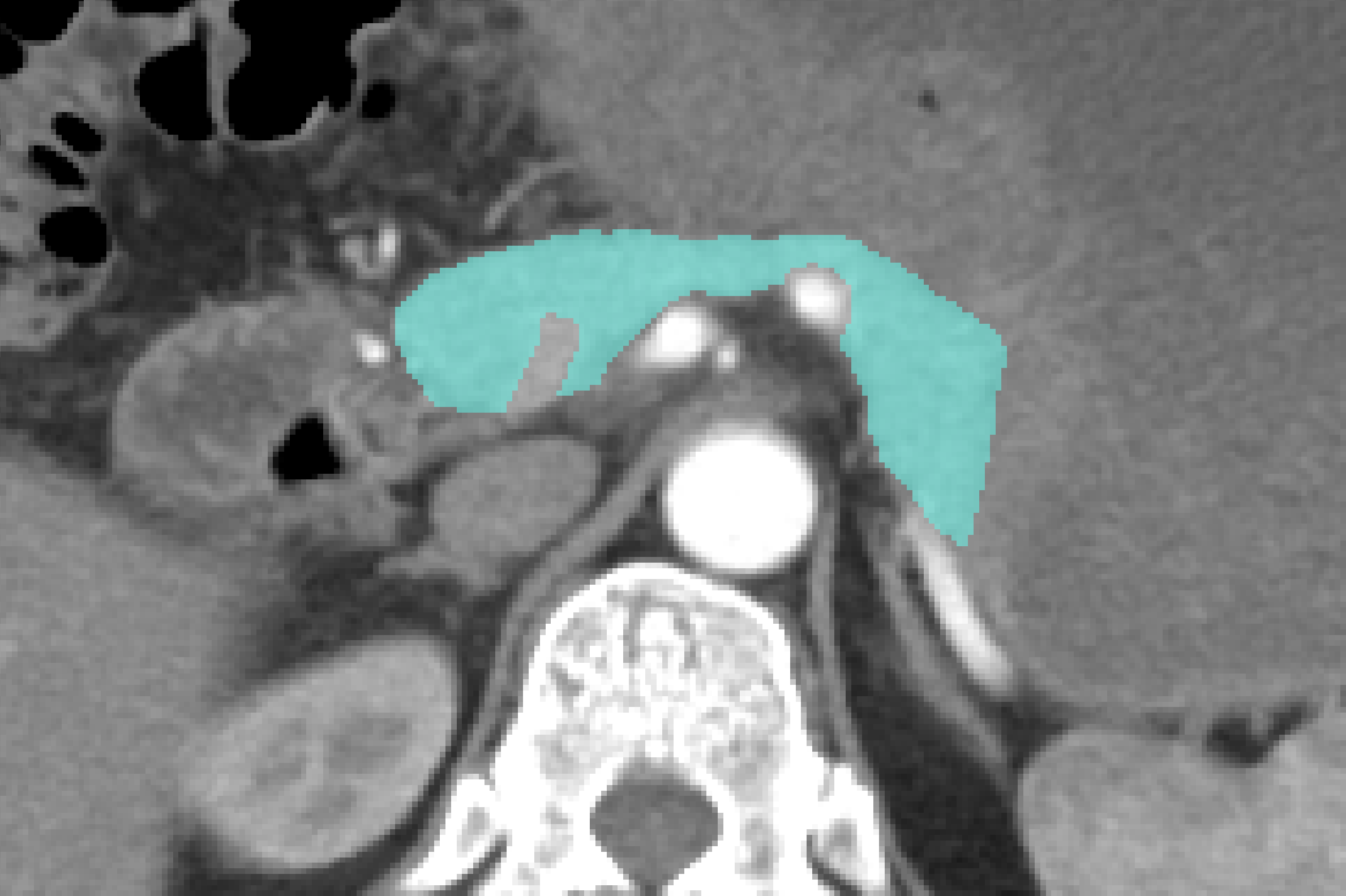}
        \caption{\small Case 1: Model 2 (\texttt{ALL$\_$45})}
    \end{subfigure}
    \hspace{0.1\textwidth}
    \begin{subfigure}[b]{0.30\textwidth}   
        \centering 
        \includegraphics[width=\textwidth]{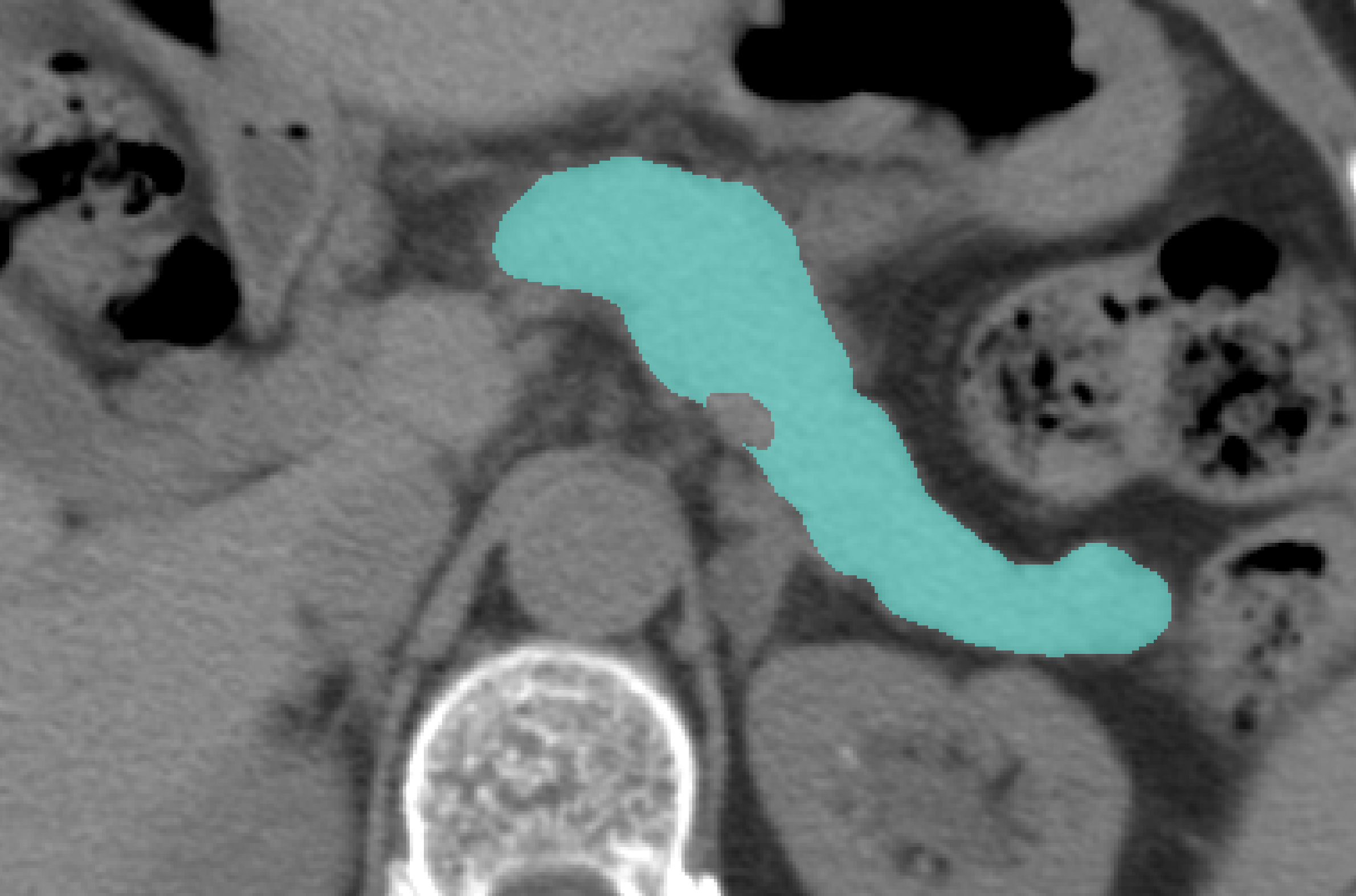}
        \caption{\small Case 2: Model 2 (\texttt{ALL$\_$45})}
    \end{subfigure}
    \smallskip
    \caption{Segmentation output for two representative scans. (a, b) Axial slices of the two scans. (c, d) Reference pancreas segmentations from the AMOS22 dataset. Segmented pancreas from (e, f) TotalSegmentator, (g, h) Model 1 (\texttt{REF$\_$8}), and (i, j) Model 2 (\texttt{ALL$\_$45}).} 
    \label{panc_seg}
\end{figure}

\end{document}